\documentclass[12pt]{article}
\usepackage{epsfig}
\def\npb#1#2#3{    {\it Nucl. Phys. }{\bf B #1} (#2) #3}
\def\plb#1#2#3{    {\it Phys. Lett. }{\bf B #1} (#2) #3}

\def\prd#1#2#3{    {\it Phys. Rev. }{\bf D #1} (#2) #3}

\def\jhep#1#2#3{   {\it JHEP  }{\bf #1} (#2) #3}

\newcommand{\ba}{\begin{array}}
\newcommand{\ea}{\end{array}}
\newcommand{\be}{\begin{equation}}
\newcommand{\ee}{\end{equation}}
\newcommand{\bea}{\begin{eqnarray}}
\newcommand{\eea}{\end{eqnarray}}
\newcommand{\beq}{\begin{equation}}
\newcommand{\eeq}{\end{equation}}

\newcommand{\Cnew}{C^{\rm new}}

\begin{document}
 
\thispagestyle{empty}
\begin{flushright}
CERN-TH/2002-263\\
SLAC-PUB-9580\\
%hep-ph/0211197
\end{flushright}

\vspace*{1.5cm}
\centerline{\Large\bf NNLL QCD Corrections to the Decay  $B \rightarrow X_s \ell^+\ell^-$}

\vspace*{2cm}
\centerline{{\large\bf  A.~Ghinculov $^{a}$, T.~Hurth $^{b,c,}$\footnote{Heisenberg Fellow},
 G.~Isidori $^{d}$, Y.-P.~Yao$^{e}$ 
}}
\bigskip

\begin{center}
{\it ${}^a$~Department of Physics and Astronomy\\ UCLA, Los Angeles CA 90095-1547, USA}\\

\vspace{0.3cm}

{\it ${}^b$~Theoretical Physics Division, CERN, CH-1211 Gen\`eve 23, Switzerland}\\

\vspace{0.3cm}

{\it ${}^c$~SLAC, Stanford University, Stanford, CA 94309, USA}\\

\vspace{0.3cm}

{\it ${}^d$~INFN, Laboratori Nazionali di Frascati, I-00044 Frascati, Italy}\\

\vspace{0.3cm}

{\it ${}^e$~Randall Laboratory of Physics\\  University of Michigan, Ann Arbor 
 MI 48109-1120, USA}
\end{center}

\centerline{\large\bf Abstract}
\begin{quote}
We briefly discuss the status of the NNLL QCD calculations 
in the inclusive rare $B$ decay $B \rightarrow X_s \ell^+ \ell^-$.
Two important ingredients, the two-loop matrix elements of the 
four quark operator ${\cal O}_2$ and the bremsstrahlung  
contributions, were quite recently finalised. 
The new contributions significantly 
improve the sensitivity of the inclusive decay $B \rightarrow X_s l^+ l^-$ 
decay in 
testing extensions of the standard model in the sector of flavour 
dynamics;  for instance  the two-loop calculation cuts 
the low-scale uncertainty in half and the bremsstrahlung calculation 
leads to a $10 \%$ shift of the position of the zero of the 
forward-backward asymmetry.~\footnote{Contribution to the 
6th International Symposium on 
Radiative Corrections Application  RADCOR 2002, Kloster Banz, Germany, 
8-13 September 2002, presented by T.H.}
\end{quote}

\newpage
\pagenumbering{arabic}
 
\section{The Decay $B \rightarrow X_s \ell^+ \ell^-$}

Inclusive rare $B$ decays like $B \rightarrow X_s \gamma$ or $B \rightarrow 
X_s \ell^+ \ell^-$ are  
very important tools to understand the nature of physics beyond the 
Standard Model (SM). The stringent bounds 
obtained from  $B \rightarrow X_s \gamma$ on 
various non-standard scenarios
(see e.g.~\cite{DGIS,Borzumati,Hurth})
are a clear example of the importance of 
theoretically clean FCNC observables
in discriminating new-physics models. 

\begin{figure}
%\begin{center}
%\leavevmode
%\epsfxsize=2cm
%\hspace{1.9cm} \epsffile[260 500 420 700]{bsll03b.eps}
\centerline{\epsfig{file=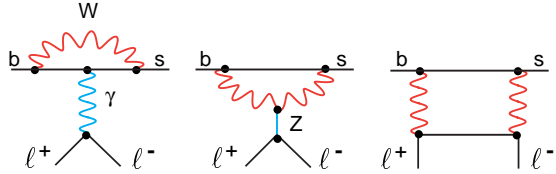,height=2.5cm}}
\caption{The decay $B \rightarrow X_s l^+ l^-$ at one-loop.}
\label{llpicture}
%\end{center}
\end{figure}

In comparison with the 
$B \rightarrow X_s \gamma$, the inclusive $ B \rightarrow X_s \ell^+\ell^-$ 
decay presents a complementary and also more complex test of the SM 
since different
contributions add to the decay rate, see fig. \ref{llpicture}.
Quite recently, BELLE announced the first measurement of 
the inclusive mode \cite{Exp}.

Generally, inclusive rare decay modes of the $B$ meson 
are theoretically  clean observables. For instance 
the decay width $\Gamma (B \rightarrow X_s \gamma)$ 
is well approximated by the partonic decay rate
$\Gamma (b \rightarrow s \gamma)$, which can be
analysed in renormalization-group-improved 
perturbation theory. Non-perturbative contributions play
only a subdominant role and can be calculated in 
a model-independent way by using the heavy-quark expansion.
However, in the decay $B \rightarrow X_s \ell^+ \ell^-$  
there are also  on-shell $c\bar{c}$ resonances.
While in the decay $B \rightarrow X_s \gamma$ (on-shell photon) 
the intermediate $\psi$ background for example, namely 
${B} \to \psi X_s$ followed by $\psi \to X' \gamma$, is 
suppressed  and can be subtracted 
from the $B \rightarrow X_s \gamma$ decay rate,
the $c\bar{c}$ resonances 
in the decay $B \rightarrow X_s l^+l^-$ (off-shell photon)
show up as large peaks in the dilepton 
invariant mass spectrum. 
These resonances can be removed by appropriate  kinematic cuts
in the invariant mass spectrum. 
In the 'perturbative windows', namely  in the low-dilepton mass 
region $s = (p_{\ell^+} + p_{\ell^-})^2 / mb^2<0.25$ and 
also in the high-dilepton mass 
region with $ 0.65<s$, 
theoretical predictions for the invariant mass spectrum
are dominated by the purely perturbative contributions, 
and theoretical precision comparable with  the one reached  
in the decay $B \rightarrow X_s \gamma$ is possible. 
In the decay $B \rightarrow X_s \ell^+ \ell^-$ kinematic observables 
such as the invariant dilepton mass spectrum and the forward-backward 
(FB) asymmetry  are particularly attractive, especially for the search for 
physics beyond the SM.
 These observables are usually normalized by the semileptonic
decay rate in order to reduce the uncertainties due 
to bottom quark mass and CKM angles. 
The normalized dilepton invariant mass spectrum 
and the FB asymmetry 
are defined as
\begin{equation}\label{decayamplitude}
R(s)= \frac{d}{d s}\Gamma(  B\to X_s\ell^+\ell^-) /
 \Gamma( B\to X_ce\bar{\nu}),
\end{equation}
\vspace{-0.3cm}
\begin{eqnarray}\label{forwardbackward}
A_{\rm FB}(s)= \frac{1}{\Gamma( B\to X_ce\bar{\nu})} 
\times \int_{-1}^1 d\cos\theta_\ell ~
 \frac{d^2 \Gamma( B\to X_s \ell^+\ell^-)}{d s ~ d\cos\theta_\ell}
\mbox{sgn}(\cos\theta_\ell), 
\end{eqnarray}
where $\theta_\ell$ is the angle between $\ell^+$ and $B$ momenta 
in the dilepton centre-of-mass frame.

The $B$ factories will soon provide statistics and resolution
needed for the measurements of $B \rightarrow X_s \ell^+\ell^-$
kinematic distributions. Precise theoretical estimates of 
the SM expectations are therefore needed in order to perform new 
significant tests of flavour physics.

\section{NNLL QCD Corrections}

Within inclusive $B$ decay modes, short-distance QCD corrections  
lead to a sizeable modification of the pure electroweak
contribution, generating large logarithms 
of the form $\alpha_s^n(m_b)$ $\log^m(m_b/M_{\rm heavy})$,
where $m \le n$ (with $n=0,1,2,...$).
A suitable framework to achieve the necessary resummations 
of these large logs is the construction of an effective low-energy 
theory with five quarks, obtained by integrating out the
heavy degrees of freedom,
\begin{equation}
\label{heffll}
H_{eff} = - \frac{4 G_{F}}{\sqrt{2}} \, \lambda_{t} \, \sum_{i=1}^{10}
C_{i}(\mu) \, {\cal O}_i(\mu) \quad .
\end{equation}
Compared with the decay $B \rightarrow X_s \gamma$,  
the effective Hamiltonian~(\ref{heffll}) contains the two 
additional operators of order $O(\alpha_{\rm em})$, 
${\cal O}_9$ and ${\cal O}_{10}$:
\begin{equation}
\begin{array}{ll}
{\cal O}_{1} = &
(\bar{s} \gamma_\mu T^a P_L c)\,  (\bar{c} \gamma^\mu T_a P_L b) \nonumber \\ 
{\cal O}_{2} = &
(\bar{s} \gamma_\mu P_L c)\,  (\bar{c} \gamma^\mu P_L b) \nonumber  \\
{{\cal O}}_{7}   = &     
  {{e}/{16\pi^2}} \, m_b(\mu) \,
 (\bar{s} \sigma^{\mu\nu} P_R b) \, F_{\mu\nu} \nonumber \\
{{\cal O}}_{8}   = &  
  {{g_s}/{16\pi^2}} \, m_b(\mu) \,
 (\bar{s} \sigma^{\mu\nu} T^a P_R b)
     \, G^a_{\mu\nu} \nonumber        \\    
{{\cal O}}_{9}   = &         
  {{e^2}/{16\pi^2}} \,
 (\bar{s} \gamma_\mu  P_L b)\, (\bar{\ell} \gamma^\mu \ell) \nonumber \\
{{\cal O}}_{10}  = &
  {{e^2}/{16\pi^2}} \,
 (\bar{s} \gamma_\mu P_L b)\,  (\bar{\ell} \gamma^\mu \gamma_5 \ell)
\end{array}
\label{heffll1}                                        
\end{equation}
The four-quark operators ${\cal O}_{3..6}$ are not given explicitly 
because of their numerically small Wilson coefficients.

Within this framework, QCD corrections are twofold: 
corrections related to the Wilson coefficients,
and those related to the matrix elements of the various operators,
both evaluated at the low-energy scale $\mu \approx  m_b$.
As the heavy fields are integrated out, the top-quark-,
$W$-, and $Z$-mass dependence is contained in the 
initial conditions of the Wilson coefficients, 
determined by a matching procedure between full and 
effective theory at the high scale. By means of RG equations, 
the $C_{i}(\mu, M_{\rm heavy})$ are then evolved at the low
scale. Finally, the QCD corrections to the matrix 
elements of the operators are evaluated at the low scale. 

Because the first large logarithm 
of the form $\log(m_b/m_W)$ arises already without 
gluons due to the mixing of the four-quark operator 
${\cal O}_2$ into ${\cal O}_9$ at one loop, 
the leading logarithms sum (LL) and the  next-to-leading logarithms sum
 (NLL) are given by 
\begin{eqnarray}
\mbox{LL} & \left[ \alpha_{\rm em}\, \log(m_b/M) \right]\,
   \alpha_s^{n}(m_b)\, \log^n(m_b/M) \nonumber \\  
\mbox{NLL} & \left[ \alpha_{\rm em}\, \log(m_b/M) \right]\,
   \alpha_s^{n+1}(m_b)\, \log^n (m_b/M) \, . \nonumber
\end{eqnarray}

The complete NLL contributions to the decay amplitude
can be found in \cite{MM,BurasMuenz}. 
Since the LL contribution to the rate turns out to be numerically 
rather small, NLL terms represent an $O(1)$ correction 
to this observable. On the other hand, since a non-vanishing 
FB asymmetry is generated by the interference of vector  
($\sim {\cal O}_{7,9}$) 
and axial-vector ($\sim {\cal O}_{10}$) leptonic currents,
the LL amplitude leads to a vanishing result and NLL terms represent 
the lowest non-trivial contribution to this observable.

For these reasons, a computation of NNLL terms
in $B \to X_s \ell^+\ell^-$ is needed 
if one aims at the same numerical accuracy as achieved 
by the NLL analysis of $B \to X_s \gamma$.
Large parts of the latter can be taken over 
and used in the NNLL calculation of
$B \to X_s \ell^+\ell^-$. However, this is not the full story.

The full computation of initial conditions of the renormalization group
equation to NNLL precision has been presented in Ref.~\cite{MISIAKBOBETH}
some time ago  - including a confirmation of the $b \to s \gamma$ NLL matching 
results of \cite{Adel}.
The inclusion of this  NNLL contribution removes the large
matching scale uncertainty (around $16 \%$) of the 
NLL calculation of the $b \to s \ell^+\ell^-$  decay rate.

Most of the NNLL contributions to the anomalous-dimension 
matrix can be derived from the NLL analysis of $b \to s \gamma$
\cite{Mikolaj}.
The missing entries are estimated to have a small 
numerical influence on the dilepton mass spectrum  \cite{MISIAKBOBETH}
and do not contribute to the FB asymmetry. 

There are two further important ingredients of the NNLL program 
which were recently calculated, namely 
the two-loop matrix elements of the four-quark operators
${\cal O}_{1,2}$ and the NNLL bremsstrahlung contributions
which will be discussed in the following sections.

In principle, a complete NNLL calculation
of the $B \rightarrow X_s \ell^+\ell^-$ rate would require also 
the calculation of  two-loop matrix element of the operator ${\cal O}_9$. 
However, its impact to the dilepton mass spectrum is also estimated to 
be very small. Similarly to the missing piece of 
the anomalous-dimension matrix, also this (scale-independent)
contribution does not enter the FB asymmetry at NNLL accuracy.

\section{Two-loop Matrix Elements of ${\cal O}_{1,2}$}

\begin{figure}
%\begin{center}
%\leavevmode
%\epsfxsize=2cm
%\hspace{1.9cm} \epsffile[260 500 420 700]{bsll03b.eps}
\centerline{\epsfig{file=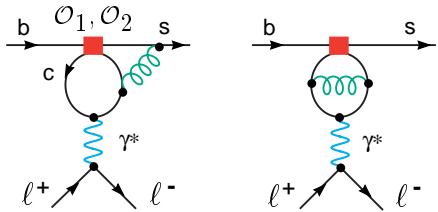,height=3.0cm}}
\caption{
Typical diagrams (finite parts) contributing to the matrix element 
of the operator ${\cal O}_{1,2}$ at the NNLL level}
\label{twoloopll}
%\end{center}
\end{figure}

Within the $B \rightarrow X_s \gamma$ calculation
at NLL, the two-loop matrix elements of the four-quark operator ${\cal O}_2$
for an on-shell photon were calculated in \cite{GHW} using Mellin-Barnes 
techniques. This calculation 
was extended in \cite{Asa1} to the case of an off-shell photon
(see fig. \ref{twoloopll})
with the help of a double Mellin-Barnes representation
which corresponds to a NNLL contribution relevant to the decay
$B \rightarrow X_s \ell^+\ell^-$.
This leads to a double expansion in the dilepton mass $s$ and 
the mass ratio $m^2_c/m^2_b$.
Thus, the validity of these analytical results given in \cite{Asa1} 
is restricted to small dilepton masses $s  < 0.25$.

An independent 
check of these results has been performed by us \cite{Adrian2}.  
Moreover, our NNLL calculation \cite{Adrian2}
is also valid for high dilepton masses for which 
the experimental methods have much higher efficiency 
compared to the one at low dilepton masses \cite{Exp}.  

In our approach  \cite{Adrian2}, the following calculational
method was used: First all diagrams were converted 
into sums of sun-set type integrals and their mass derivatives,
$$ \int d^{n}p\,d^{n}q\, 
       \frac{p^{\mu_1} \ldots p^{\mu_i} q^{\mu_{i+1}} \ldots q^{\mu_j}}{
             ((p+k)^{2}+m_{1}^2) 
             (q^{2}+m_{2}^{2})
             (r^{2}+m_{3}^{2})}$$
where $r=p+q$. The effective masses $m_{1,2,3}^2$ and the 
effective momentum $k$ are 
polynomial  functions of physical masses, external kinematics, and
Feynman parameters associated with the diagrams. Integrations over the 
Feynman parameters are understood. After the internal momenta $p$ and $q$ 
are integrated over, one can show that the results can be spanned by a 
finite set of ten scalar kernels, which 
are functions of $m_{1,2,3}^2$ and $k^2$
in one-dimensional integral representations, multiplied by tensors 
made of the metric and $k$. The Feynman parameters are extended
into the complex plane to effect rapid numerical convergence.
In principle, possible IR singularities have to be isolated and 
substracted.  However, 
an important simplification within this specific calculation  
is given by the fact
that  all relevant two-loop diagrams are IR finite. 

\begin{figure}
%\begin{center}
%\leavevmode
%\epsfxsize=2cm
\centerline{\epsfig{figure=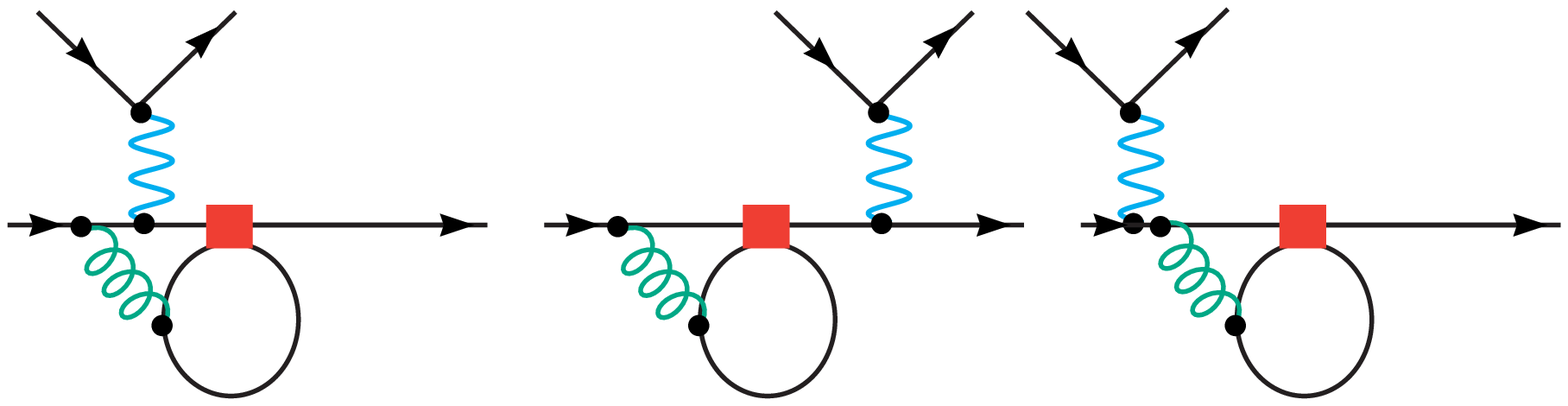,height=2.4cm}}
\centerline{\epsfig{figure=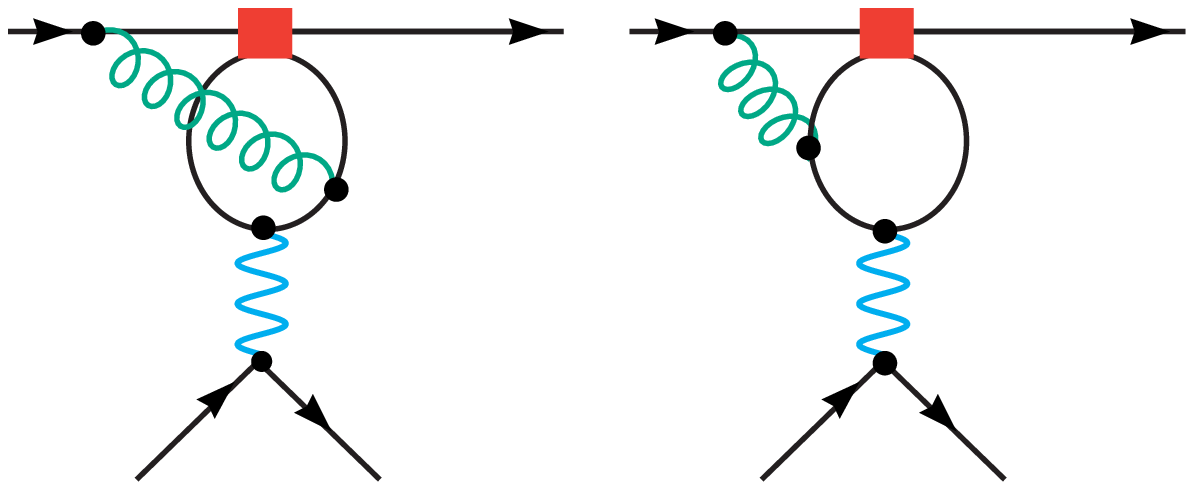,height=2.4cm}}
\caption{Two gauge-invariant subset, without (up) and with (down) 
a nontrivial $c \bar c$ threshold.}
%without $c \bar c$ threshold.}
\label{twoloopll11}
%\end{center}
\end{figure}

It is clear  that a nontrivial $c \bar c $ threshold behaviour at the NNLL
level cannot be reproduced by the expansion method used in \cite{Asa1}.
This easily explains our results of the comparison 
of the two independent calculations, \cite{Asa1} and \cite{Adrian2}, 
within the low-s-region.
The Mellin-Barnes expansion of the gauge-invariant subset shown in fig.
\ref{twoloopll11} are in excellent agreement with our numerical results.
We found that the expanded results given in \cite{Asa1} 
are even  valid beyond the claimed validity range $s < 0.25$.  
In contrast, this is not true for the second gauge-invariant subset given 
in fig. \ref{twoloopll11} because of the nontrivial 
$c \bar c$ threshold in that case.
In \cite{Asa1} it was already shown  that in the 
low-dilepton mass region 
these NNLL  contributions reduce the perturbative uncertainty (due 
the low-scale dependence) 
from $\pm 13\%$ down to $\pm 6.5\%$ and also the central value
is changed significantly, $\sim 14 \%$. 

There is no additional problem due to the charm mass renormalisation
scheme ambiguity within the decay $B \rightarrow X_s \ell^+ \ell^-$ 
because the charm dependence starts already at one-loop in contrast 
to the case of the decay $B \rightarrow X_s \gamma$. 
The charm dependence itself leads to a $\sim 7 \%$ uncertainty. 
These small uncertainties in the inclusive mode should be compared 
with the ones of the correspondung exclusive mode 
$B \rightarrow K^* \mu^+ \mu^-$  given in \cite{Balletal};
$\Delta BR 
= (^{+26}_{-17}, \pm 6, ^{+6}_{-4}, ^{-0.7}_{+0.4}, 
\pm 2)\% $.  The first dominating error represents the hadronic 
uncertainty due to the 
formfactors.

A phenomenolgical  NNLL analysis including the high dilepton mass region 
will be  presented in \cite{Adrian2}.

\section{Bremsstrahlung Contributions}

\begin{figure}
%\begin{center}
%\leavevmode
%\epsfxsize=2cm
%\hspace{1.9cm} \epsffile[260 500 420 700]{bsll03b.eps}
\centerline{\epsfig{file=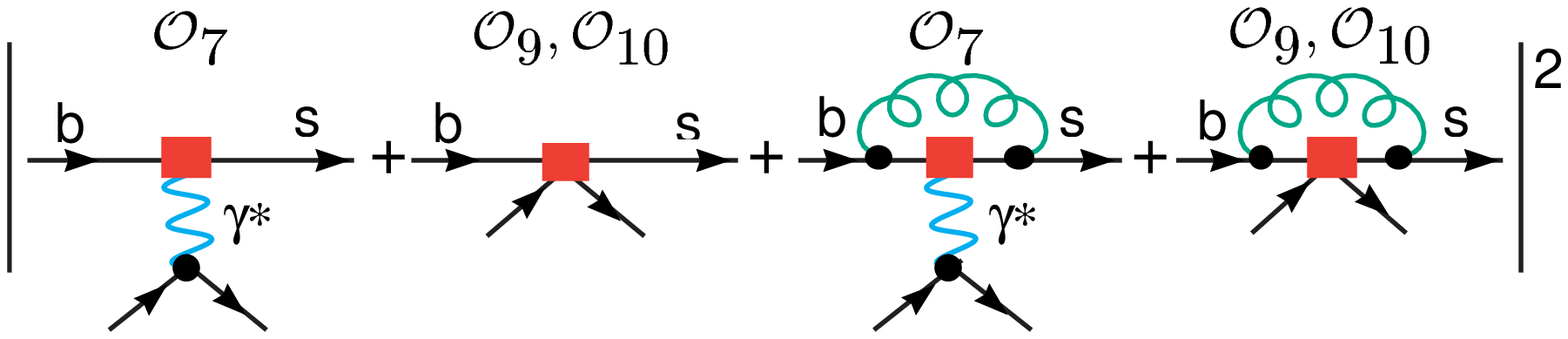,height=2.3cm}}
\centerline{\epsfig{file=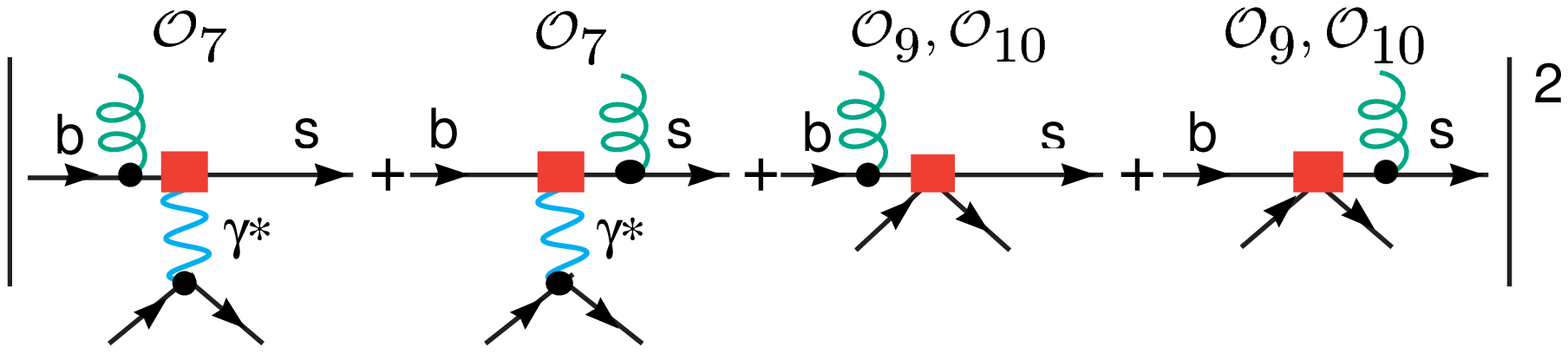,height=2.3cm}}
\caption{Virtual (up) and real (down) QCD corrections.}
% generating the 
%terms $\tau_i$ and $\sigma_i$ in Eqs.~(\ref{effmod}).}
\label{bsll03}
%\end{center}
\end{figure}

The NNLL bremsstrahlung contributions were also recently calculated 
for the dilepton mass spectrum (symmetric part) in 
\cite{Asa3,Adrian1} and  for the FB asymmetry in \cite{Adrian1,Asa2},
see fig. \ref{bsll03}.

In  \cite{Adrian1} we have separated the bremsstrahlung
corrections into universal and nonuniversal pieces.
We defined the terms $\sigma_{7,9}(s)$ 
which take into account {\em universal} $O(\alpha_s)$ bremsstrahlung 
corrections as 
$O(\alpha_s)$ corrections of the effective Wilson coefficients
by demanding that the remaining (finite) 
non-universal bremsstrahlung corrections of the rate 
(encoded in the functions $\tau_{77}$ and $\tau_{99}$)
vanish in the limit $s \rightarrow 1$.
Then the universal contributions
take into account the truly soft component of the radiation, 
which diverges at the $s\to 1$ boundary of the phase space. 
This is because in the $s \rightarrow 1$ limit only 
the soft component of the radiation survives and, according 
to Low's theorem, the latter gives rise to a correction proportional 
to the tree-level matrix element.
In \cite{Adrian1} we then found that {\it all} nonuniversal bremsstrahlung
corrections $\tau_i$, to the rate and to the FB asymmetry, 
are rather small all over the phase space, and particularly
for large values of $s$ ($|\tau_i(s)|< 0.5$ for $s>0.3$),  in comparison
with the dominating universal corrections.

In the case of the forward--backward asymmetry
there are ambiguities arising for $d\not=4$
in the definition of $\gamma_5$ but 
in the case of the decay rate, the problematic $\gamma_5$ 
contribution vanishes because of the $p_1 \leftrightarrow p_2$ permutation 
symmetry of the leptonic phase space. 
To circumvent this problem, we employed the following hybrid 
regularization scheme \cite{Adrian1}: 
the Dirac algebra of IR-divergent pieces 
is strictly treated in four dimensions (dimensional reduction), while 
the virtual UV-divergent pieces,  
which do not involve any $\gamma_5$ ambiguity, are 
still computed in na\"\i ve dimensional regularization.
For example, the hybrid wave function renormalization constant for a massless
quark is given by
$ Z_\psi^{m=0} =  1 - \frac{\alpha_s}{4\pi} \frac{4}{3} 
 \left( \frac{1}{\epsilon_{\rm UV}} - \frac{1}{\epsilon_{\rm IR}}  - 1 \right)$. At this level of the perturbative expansion, 
this hybrid regularization scheme is gauge invariant.
Using this scheme we were able to explicitly verify the 
cancellation of IR divergences in our bremsstrahlung calculation
\cite{Adrian1}.
%A different strategy has been adopted in \cite{Asa2} where, 
%anticipating the cancellation of IR-divergences, the $\gamma_5$ 
%ambiguity was circumvented by calculating only finite 
%bremsstrahlung contributions. 
However, anticipating the cancellation of IR-divergences, one also can 
circumvent the $\gamma_5$ ambiguity by calculating only finite 
bremsstrahlung contributions \cite{Asa2}. 

Let us finally discuss the phenomenological impact of these 
bremsstrahlung calculations - focusing on the  
position of the zero of the FB asymmetry. 
This quantity, defined by $A_{\rm FB}(s_0)=0$, 
is particularly interesting to determine relative sign and 
magnitude of the Wilson coefficients $C_7$ and $C_9$ and it is
therefore extremely sensitive to possible new physics effects.

\begin{figure}%figffunction
%\begin{center}
%\leavevmode
%\epsfxsize=2cm
\centerline{\epsfig{file=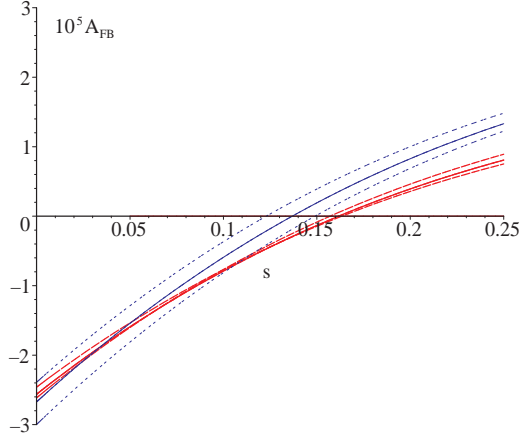,height=6.8cm}}
\caption{Comparison between NNLL and  NLL results for 
$A_{\rm FB}(s)$ in the low $s$ region. 
The three thick (red) lines are the NNLL  
predictions for $\mu=5$~GeV (full),
and $\mu=2.5$ and 10 GeV (dashed); the dotted (blue) curves 
are the corresponding NLL results. All curves for $m_c/m_b=0.29$.}
\label{fig:AFB}
%\end{center}
\end{figure}

The NLL result $s^{\rm NLL}_0 = 0.14 \pm 0.02~$
where the error is determined by the scale dependence 
($2.5~\rm{ GeV} \leq \mu \leq 10$~GeV) is now modified by 
the  discussed NNLL contributions to (see fig. \ref{fig:AFB})
\begin{equation}
s^{\rm NNLL}_0 = 0.162 \pm 0.008~.
\label{eq:s0NLO}
\end{equation}
In this case the variation of the result induced by the scale dependence is 
accidentally very small (about $\pm 1\%$ for $2.5~\rm{ GeV} \leq \mu \leq 10$~GeV) and cannot be regarded as a good estimate of missing higher-order effects. 
Taking into account the separate scale variation of both Wilson 
coefficients $\Cnew_9$ and  $\Cnew_7$,
and the charm-mass dependence, we estimate a conservative overall error
on $s_0$ of about $5 \%$ \cite{Adrian1}. 
In this $s$-region the nonperturbative  
$1/m_b^2$ and $1/m_c^2$ corrections to $A_{\rm FB}(s)$ 
are very small and also under control \cite{Falk,Handoko,Gino,Christian}.
Summing up, the zero of the FB asymmetry in the inclusive mode turns out 
to be one of the most sensitive tests for new physics beyond the SM.  

\end{document}